\documentclass[aps,prl,reprint,nofootinbib,superscriptaddress]{revtex4-2}

\usepackage[T1]{fontenc}
\usepackage[utf8]{inputenc}
\usepackage{amsmath,amssymb,mathtools}
\usepackage{graphicx}
\usepackage[dvipsnames]{xcolor}
\usepackage{hyperref}
\usepackage{CJKutf8}

\usepackage{tikz}
\usetikzlibrary{decorations.markings}

\usepackage{graphicx}
\usepackage{dcolumn}
\usepackage{enumitem}
\usepackage{bm}
\usepackage{float}
\usepackage[many]{tcolorbox}
\usepackage{environ}
\usepackage{physics}
\usepackage{tabularx}
\usepackage{diagbox}
\usepackage{subcaption}
\usepackage{multirow}
\AtBeginDocument{\setcounter{secnumdepth}{4}}

\graphicspath{{fig/}}

\newcommand{\Lagr}{\mathcal{L}}
\newcommand{\piphi}{\mathrm{NLSM}{+}\phi^3}

\begin{document}
	
	\begin{CJK*}{UTF8}{gbsn}
		
		\title{NLSM amplitudes from a quartic two-derivative theory}
		
		\author{Qu Cao(曹趣)}
		\email{caoqu@westlake.edu.cn}
		\affiliation{Department of Physics, School of Science, Westlake University, Hangzhou 310030, China}
		\affiliation{Institute of Natural Sciences, Westlake Institute for Advanced Study, Hangzhou 310024, China}

		\author{Zhen-qi Han(韩振琦)}
		\email{hanzhenqi24@mails.ucas.ac.cn}
		\affiliation{School of Fundamental Physics and Mathematical Sciences, Hangzhou Institute for Advanced Study, UCAS and ICTP-AP, Hangzhou 310024, China}
		\affiliation{Institute of Theoretical Physics, Chinese Academy of Sciences, Beijing 100190, China}
		\affiliation{School of Physical Sciences, University of Chinese Academy of Sciences, Beijing 100049, China}
		
		\author{Fan Zhu(朱凡)}
		\email{zhufan25@gscaep.ac.cn}
		\affiliation{Graduate School of China Academy of Engineering Physics, Beijing 100193, China}
		
		\begin{abstract}
			We revisit the well-known nonlinear sigma model (NLSM), an effective field theory describing the scattering of $\mathrm{SU}(N)$ Goldstone bosons and characterized by an infinite tower of two-derivative interactions. We introduce a local scalar Lagrangian involving two scalar fields $\psi^\pm$ carrying opposite ``polarities'', whose interacting part consists of a single polynomial quartic two-derivative operator, and prove that it reproduces planar NLSM amplitudes at all loop orders. This quartic two-derivative formulation reveals a previously hidden simplicity of the NLSM: its Feynman rules involve only a single quartic interaction vertex, making the Adler zero and the leading double-soft factor manifest at all loop orders on generalized cuts and significantly improving the efficiency of high-multiplicity computations. We also suggest a possible analogous description of the $\piphi$ theory in terms of a finite tower of local interactions.
		\end{abstract}
		\maketitle
	\end{CJK*}

	\section{Introduction}
	
	The nonlinear sigma model (NLSM) is the universal two-derivative effective field theory of Goldstone bosons~\cite{Coleman:1969sm,Callan:1969sn,Kampf:2013vha}. Its Lagrangian is
	\begin{equation}
		\mathcal{L}_{\pi}^{\rm NLSM}
		=
		\frac{1}{2}\Tr\!\left(\partial_\mu U\,\partial^\mu U^\dagger\right),
		\quad
		U(x)\in \mathrm{SU}(N),
		\label{eq:intro-nlsm-lagrangian}
	\end{equation}
	where $U$ is the Goldstone matrix\footnote{In this Letter, we set the pion decay constant to $F_\pi=1$.}. Expanding $U$ in pion fields generates an infinite tower of two-derivative interactions, and the flavor-dressed amplitudes decompose into trace structures and flavor-ordered partial amplitudes.
	\begin{equation}
		\mathcal{A}^{\pi}_{n}
		=
		\sum_{\sigma\in S_n/\mathbb{Z}_n}
		\Tr(T^{a_{\sigma(1)}}\cdots T^{a_{\sigma(n)}})\,
		A^{\pi}_{n}(\sigma)\,.
		\label{eq:intro-flavor-decomp}
	\end{equation}
	In this Letter we focus on planar partial amplitudes $A_n^\pi$ with the canonical ordering $(1,2,\ldots,n)$. They obey the Adler zero: an on-shell amplitude vanishes when any external pion is taken soft~\cite{Adler:1964um,Susskind:1970gf}. In the conventional perturbative expansion, however, this property is not manifest diagram by diagram but emerges from nontrivial cancellations among Feynman diagrams~\cite{Du:2015esa}.
	
	Modern amplitude methods reveal a much simpler on-shell structure. Berends--Giele recursion~\cite{Kampf:2013vha,ChenDu:2013AmpRel,LowYin:2017Ward,Mizera:2018jbh}, soft recursion~\cite{Cheung:2014dqa,Cheung:2015ota,Rodina:2016jyz,LowYin:2019SoftBootstrap}, flavor--kinematics~\cite{Bern:2008qj,Cheung:2016fjp,DuFu:2016BCJ,Edison:2023ulf}, double-copy~\cite{Bern:2010ue,Cheung:2017yef,Cheung:2017ems}, geometric duality~\cite{Cheung:2021yog,Cheung:2022vnd,Brauner:2025rzv,Helset:2024vle}, Abelian $Z$-theory~\cite{Carrasco:2016ldy}, and the CHY formalism~\cite{Cachazo:2013hca,Cachazo:2013iea,Cachazo:2014xea} provide increasingly compact descriptions of NLSM amplitudes. More recently, planar NLSM amplitudes were related to cubic scalar amplitudes through special kinematic shifts and hidden-zero structures~\cite{Arkani-Hamed:2023swr,Arkani-Hamed:2024nhp}. These developments suggest that the complexity of the standard Feynman rules is largely an artifact of the chosen variables.
	
	Motivated by this, we reformulate the planar NLSM as a local quartic two-derivative scalar theory with two auxiliary fields $\psi^\pm$ and a single quartic interaction.  The $\pm$ labels are bookkeeping polarities assigned around a fixed planar ordering, rather than physical charges.  Summing the quartic diagrams exactly reproduces the NLSM amplitudes. 
	This formulation makes the Adler zero and the leading double-soft factor~\cite{Kampf:2013vha, Cachazo:2015ksa,Low:2015ogb} manifest at all loop orders on generalized cuts~\cite{Bern:1994zx,Britto:2004GenUnitarity,BernHuang:2011GenUnitarity}, gives a compact tree-level organization of the leading single-soft coefficient~\cite{Cachazo:2016njl}, and provides an efficient high-multiplicity expansion.  We also attempt to extend the construction to mixed NLSM$+\phi^3$ amplitudes.
	
	\section{The Quartic Two-Derivative Scalar Theory}\label{sec:quartic}
	Our quartic two-derivative scalar theory is defined in terms of two auxiliary fields of opposite ``polarities'', $\psi^+$ and $\psi^-$.  We call this theory QTDS, for ``quartic two-derivative scalar''.  Planar flavor-ordered partial amplitudes of the NLSM will emerge from its local Feynman rules, and the Lagrangian is defined as
	\begin{equation}
		\Lagr^{\rm QTDS}_{\psi}
		{=}
		\Tr(\partial_\mu\psi^+\,\partial^\mu\psi^-)
		{+}
		2\Tr(\partial^\mu\psi^+\psi^-\partial_\mu\psi^+\psi^-).
		\label{eq:qtds-action}
	\end{equation}
	The trace notation only keeps track of the cyclic ordering of the flavor generators, while the overall flavor factors are stripped off.  The kinetic term couples $\psi^+$ only to $\psi^-$, and therefore the only propagator is
	$\langle\psi^+(k)\psi^-(-k)\rangle{\sim}1/k^2$.  Throughout this paper, all momenta are taken to be outgoing, and the only interaction term gives the quartic vertex
	\begin{equation}
		\boxed{V^{(4)}(\psi_1^+\psi_2^-\psi_3^+\psi_4^-)=-2 k_1{\cdot} k_3\,.}
		\label{eq:quarticrule}
	\end{equation}
	We assign opposite polarities to adjacent external legs, while every internal propagator connects fields of opposite polarities.  We should emphasize that QTDS is not introduced as a conventional field redefinition of the NLSM Lagrangian; rather, it is an auxiliary local theory whose planar Feynman rules reproduce the same flavor-ordered amplitudes.  Our main claim is that the tree amplitudes and loop integrands of the NLSM are reproduced by the QTDS Feynman rules~\eqref{eq:quarticrule} in the following sense, with $n$ even:
	\begin{equation}
		\begin{cases}
			A^{\pi}_{n}(\pi_1\pi_2\ldots\pi_{n})
			=A^{\psi}_{n}(\psi_1^{\pm}\psi_2^{\mp}\ldots\psi_{n}^{\mp})\\[3pt]
			
			I^{\pi,(L)}_{n}(\pi_1\pi_2\ldots\pi_{n})
			\cong I^{\psi,(L)}_{n}(\psi_1^{\pm}\psi_2^{\mp}\ldots\psi_{n}^{\mp})
		\end{cases}\,,
		\label{eq:pipsi}
	\end{equation}
	Here $A^{\pi/\psi}$ denotes tree amplitudes, $I^{\pi/\psi}$ denotes loop integrands, and $\cong$ denotes equality up to scaleless integrands, which vanish after integration. 
	In Appendix~\ref{app:proof}, we give a diagrammatic proof of~\eqref{eq:pipsi} at the level of generalized unitarity cuts, using planar variables~\cite{Arkani-Hamed:2017mur,Arkani-Hamed:2024tzl}.  For simplicity, we always consider the canonical ordering.  At tree level, flipping all polarities gives the same amplitude.  At loop level, the two flipped assignments need not give identical integrand representatives; starting at sufficiently high loop order they can differ by scaleless terms, and are therefore equivalent only after integration.
	
\begin{figure}[htbp]
		\centering
		\includegraphics[width=1\linewidth]{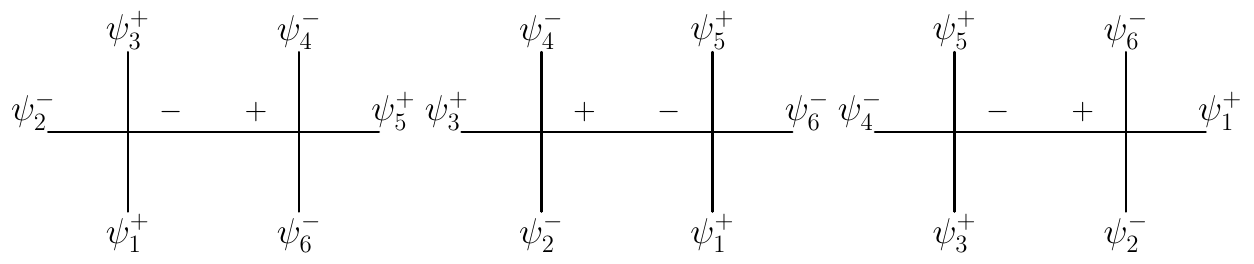}
		\captionsetup{justification=raggedright,singlelinecheck=false}
		\caption{The three quartic diagrams contributing to the six-point tree amplitude $A^{\psi}_6(\psi_1^+\psi_2^-\psi_3^+\psi_4^-\psi_5^+\psi_6^-)$.}
		\label{fig:6pexp}
	\end{figure}

	The first nontrivial example is the six-point tree amplitude shown in Fig.\ref{fig:6pexp}.  The explicit result is
	\begin{equation}
		A^{\psi}_6{=}
		\frac{4\,k_1{{\cdot}} k_3\, k_5{{\cdot}} k_{1,2,3}}{s_{1,2,3}}
		{+}\frac{4\,k_1{{\cdot}} k_5\, k_3{{\cdot}} k_{5,6,1}}{s_{2,3,4}}
		{+}\frac{4\,k_3{{\cdot}} k_5\, k_1{{\cdot}} k_{3,4,5}}{s_{3,4,5}}\,,
		\label{eq:A6}
	\end{equation}
	where $k_{a,b,c}{=}k_a{+}k_b{+}k_c$ and $s_{a,b,c}{=}k_{a,b,c}^2$ are the Mandelstam variables.  One can check that~\eqref{eq:A6} is exactly the NLSM amplitude $A^\pi_6$.
	
	\noindent{\bf Efficiency}---The gain of the QTDS formulation is already visible from diagram counting.  At fixed multiplicity, QTDS involves only quartic trees, fewer than both cubic trees and the usual even-valent NLSM trees; see Table~\ref{tab:qtds-counting} in Appendix~\ref{app:proof} for a detailed comparison.  Together with the simple numerator in Eq.~\eqref{eq:quarticrule}, this gives an efficient diagrammatic expansion for high-multiplicity tree amplitudes. The ancillary \texttt{Mathematica} code evaluates the 20-point tree amplitude in a few minutes.
	
	\section{Possible Extension to \texorpdfstring{NLSM$+\phi^3$}{NLSM+phi3}}
	\label{sec:mixed}
	
	Mixed NLSM$+\phi^3$ amplitudes provide a natural next test of the quartic reformulation.  They appear in several amplitude constructions: they admit compact CHY representations~\cite{Cachazo:2016njl}, arise from semi-abelian $Z$-theory in the low-energy expansion~\cite{Carrasco:2016ygv}, and encode the first nonzero coefficient in the single-soft expansion of pure NLSM amplitudes~\cite{Cachazo:2016njl,LowYin:2017Ward,LowYin:2018,Dong:2021qai}. In this section, we focus on the single-trace theory $\mathrm{NLSM}{+}\Tr(\phi^3)$, where the scalar $\phi$ shares the planar ordering with the pions and has a cubic self-interaction $V^{(3)}(\phi_1\phi_2\phi_3){=}1$.  The closely related NLSM plus bi-adjoint $\phi^3$ theory introduces a second color ordering for the scalars~\cite{Cachazo:2016njl,Carrasco:2016ygv}; this changes the allowed planar diagrams but not the local pion--scalar vertex patterns discussed below.
	
	As in the pure NLSM sector, mixed amplitudes vanish unless the number of external pions is even.  When only two external $\phi$ scalars are present, the mixed partial amplitude is equivalent to the corresponding pure-pion partial amplitude with the same ordering~\cite{Cachazo:2016njl,LowYin:2017Ward,Dong:2021qai}.
	
	We keep the same scalar fields $\psi^\pm$ defined in the QTDS, while the scalar $\phi$ carries no polarity label.  Matching the four-point mixed amplitudes with two pions and two scalars suggests five quartic $\psi$--$\phi$ vertices. 
	Their numerator factors are inherited from the same two-derivative structure as the pure QTDS vertex.  The explicit vertex rules are
	\begin{equation}
		\begin{cases}
			V^{(4)}(\psi^+_1\psi^-_2\phi_3\phi_4)={-}2k_1{\cdot} k_3\,,\\[3pt]
			V^{(4)}(\psi^+_1\phi_2\phi_3\psi^-_4)={-}2k_1{\cdot} k_3\,,\\[3pt]
			V^{(4)}(\psi^+_1\phi_2\psi^-_3\phi_4)={-}2k_1{\cdot} k_3\,,\\[3pt]
			V^{(4)}(\psi^+_1\phi_2\psi^+_3\phi_4)={-}2k_1{\cdot} k_3\,,\\[3pt]
			V^{(4)}(\phi_1\psi^-_2\phi_3\psi^-_4)={-}2k_1{\cdot} k_3\,.
		\end{cases}
		\label{eq:mix4rule}
	\end{equation}
	The idea is to supplement the pure QTDS rules by local $\psi$--$\phi$ vertices, while assigning polarities only to the $\psi$ fields.  In contrast to the pure NLSM sector, however, the polarities of the external $\psi$ fields cannot be chosen freely if one wants to reproduce the correct mixed amplitudes.  As in the pure QTDS construction, adjacent $\psi$ fields are still required to carry opposite polarities.  Some allowed low-point patterns are summarized in Table~\ref{tab:mixed-polarities}, all mixed NLSM tree amplitudes up to seven points are generated by the corresponding Feynman diagrams built from the local rules described above.
	\begin{table}[htbp]
		\centering
		\footnotesize
		\setlength{\tabcolsep}{4pt}
		\renewcommand{\arraystretch}{1.65}
		\begin{tabular}{@{}c c@{}}
			\hline\hline
			$n$ & Polarity patterns\\
			\hline
			$5$ & \makebox[0.85\linewidth][c]{$\psi^\pm\psi^\mp\phi\phi\phi$, $\psi^+\phi\psi^+\phi\phi$}\\
			
			$6$ & \makebox[0.85\linewidth][c]{$\psi^\pm\psi^\mp\phi\phi\phi\phi$, $\psi^+\phi\psi^+\phi\phi\phi$, $\psi^+\phi\phi\psi^+\phi\phi$}\\[3pt]
			
			$7$ & \begin{minipage}{0.85\linewidth}
				\centering
				$\psi^\pm\psi^\mp\psi^\pm\psi^\mp\phi\phi\phi$, $\psi^\pm\psi^\mp\phi\psi^+\phi\psi^+\phi$\\
				$\psi^+\psi^-\phi\psi^-\psi^+\phi\phi$, $\psi^+\psi^-\psi^+\phi\psi^+\phi\phi$, $\psi^+\psi^-\psi^+\phi\phi\psi^+\phi$
			\end{minipage}\\
			\hline\hline
		\end{tabular}
		\captionsetup{justification=raggedright,singlelinecheck=false}
		\caption{Some allowed low-point polarity patterns for mixed amplitudes.}
		\label{tab:mixed-polarities}
	\end{table}
	
\begin{figure}[htbp]
		\centering
		\includegraphics[scale=0.48]{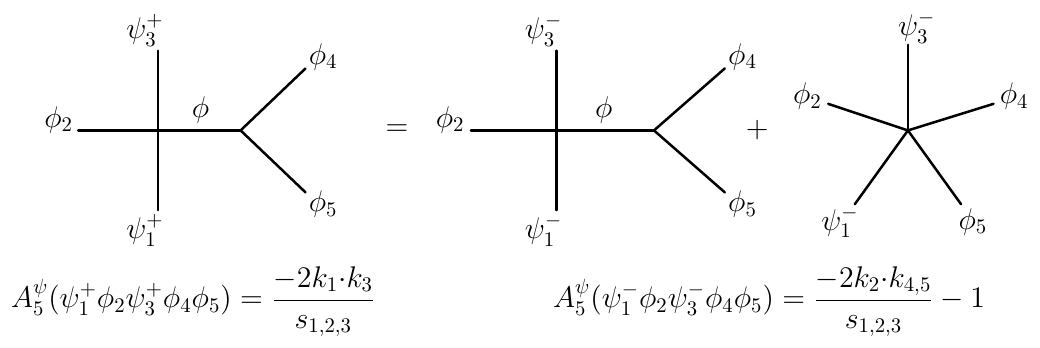}
		\captionsetup{justification=raggedright,singlelinecheck=false}
		\caption{The mixed amplitude for ordering $\psi^+_1\phi_2\psi^+_3\phi_4\phi_5$ and its globally flipped polarity assignment.}
		\label{fig:mix-flip-example}
	\end{figure}

	However, the quartic mixed vertices~\eqref{eq:mix4rule} are not sufficient at higher multiplicity.  Starting at $n{\geq}8$, the amplitudes generated by the quartic vertices described above fail to reproduce some mixed amplitudes, for example the ordering $\psi^+_1\psi^-_2\phi_3\phi_4\psi^-_5\psi^+_6\phi_7\phi_8$.  In this case the correct mixed amplitude is recovered after introducing the penta-vertex
	\begin{equation}
		V^{(5)}(\psi^-\phi\psi^-\phi\phi)=-1\,,
		\label{eq:pentavertex}
	\end{equation}
	which generates additional Feynman diagrams and supplies the missing contribution.  
	After including the five-point vertex in Eq.~\eqref{eq:pentavertex}, all polarity patterns listed in Table~\ref{tab:mixed-polarities} can also be globally flipped, and the resulting Feynman rules still reproduce the correct mixed amplitudes.  A five-point example is shown in Fig.~\ref{fig:mix-flip-example}: the original assignment is generated by one diagram, while the flipped assignment requires the additional diagram with a penta-vertex, and the two computations give the same mixed amplitude.  We have checked this statement for all mixed amplitudes up to eleven points in the following class of orderings: 
	\begin{quote}
		\it the external $\psi$ legs form at most two consecutive $\psi$-blocks of length greater than one, separated by $\phi$ legs.
	\end{quote}
	For all such $\psi$--$\phi$ distributions, the vertices described above reproduce the corresponding NLSM$+\phi^3$ mixed amplitudes for suitable polarity assignments.  Further details and examples are given in Appendix~\ref{app:mixed}.
	This shows that, unlike the pure NLSM case, the mixed extension is sensitive not only to vertex valency but also to the global species and polarity pattern.  Whether all NLSM$+\phi^3$ mixed amplitudes can be described by a finite tower of local vertices remains an open question.
	
	\section{Soft Behaviors}\label{sec:soft}
	Soft behavior is one of the most characteristic signatures of the NLSM.  It encodes the nonlinear Goldstone symmetry directly at the level of scattering amplitudes, and provides a stringent check on any representation of the theory~\cite{Adler:1964um,Susskind:1970gf,Kampf:2013vha,Cachazo:2015ksa,Du:2015esa,Low:2015ogb,LowYin:2017Ward,LowYin:2019SoftBootstrap,Rodina:2021isd}.  In this section we show that the quartic two-derivative scalar rules make the Adler zero and the leading double-soft factor transparent at all loop orders on generalized cuts.  We also explain how the same rules organize the leading coefficient in the tree-level single-soft expansion.
	
	\subsection{Single-soft behaviors}
	For a single soft pion, $k_i{=}\tau p$ with $\tau{\to}0$, the Adler zero states that the order-$\tau^0$ term vanishes, $\lim_{k_i\to0}A^\pi_n=0$.  
	At tree level, the Adler zero, together with locality and consistent factorization, fixes the NLSM amplitudes uniquely~\cite{Arkani-Hamed:2016rak,Rodina:2016jyz,Rodina:2018pcb,Cheung:2014dqa,Cheung:2015ota}, and is closely related to gauge invariance of gauge particles~\cite{Dong:2024klq}.
	
	In the standard Feynman-diagram expansion the zero is hidden: diagrams with different even-valent contact vertices cancel only after the full sum is taken~\cite{Du:2015esa}.  In the quartic rules the same statement is reorganized locally.  For a fixed polarity assignment, every external leg in one polarity sector appears through an explicit momentum factor.  At tree level, this can be written schematically as
	\begin{equation}
		\begin{aligned}
			&A^{\psi}_{n}(\psi_1^+ \psi_2^-\ldots \psi_{n}^-)\\
			=&\,k_1^{\mu_1}k_3^{\mu_3}{\cdots} k_{n-1}^{\mu_{n-1}}\,
			J^{\psi}_{\mu_1\mu_3\ldots\mu_{n-1}}
			(\psi_1^+ \psi_2^-\ldots \psi_{n}^-)\,;
			\\[3pt]
			&A^{\psi}_{n}(\psi_1^- \psi_2^+\ldots \psi_{n}^+)\\
			=&\,k_2^{\mu_2}k_4^{\mu_4}{\cdots} k_{n}^{\mu_{n}}\,
			J^{\psi}_{\mu_2\mu_4\ldots\mu_{n}}
			(\psi_1^- \psi_2^+\ldots \psi_{n}^+)\,.
		\end{aligned}
		\label{eq:psiadlerzero1}
	\end{equation}
	Here $J^\psi$ denotes the on-shell current obtained after stripping off the momenta carried by one polarity sector.  Thus one polarity assignment makes the Adler zeros of the odd legs manifest, while the globally flipped assignment makes those of the even legs manifest.  Since the two tree amplitudes are identical, all tree-level Adler zeros can be made manifest in this representation.
	
	At loop level the situation is subtle. Consistently with Refs.~\cite{Bartsch:2022pyi,Bartsch:2024ofb}, no single integrand representative satisfying generalized unitarity cuts is expected, starting at two loops, to make all Adler zeros manifest at once. Instead, the two globally flipped polarity assignments provide two integrand representatives with the same current-factorized form as in Eq.~\eqref{eq:psiadlerzero1}.  Each assignment therefore makes a complementary half of the Adler zeros manifest.
	For $L{\geq}2$ and $n{\geq}4$, these two representatives are generally not identical at the integrand level,
	\begin{equation}
		I^{\psi,(L\geq2)}_{n\geq4}(\psi_1^+ \psi_2^-\ldots \psi_{n}^-)
		\cong
		I^{\psi,(L\geq2)}_{n\geq4}(\psi_1^- \psi_2^+\ldots \psi_{n}^+)\,,
	\end{equation}
	although their difference is a sum of scaleless terms that vanish after integration. Thus, at loop level, QTDS does not provide a single representative that makes all Adler zeros manifest; instead, the two polarity choices manifest complementary halves and become equivalent after integration.  In Appendix~\ref{app:proof}, we also illustrate how $I^{\psi,(L)}$ manifests the Adler zeros in planar variables through the algebraic soft limits in Ref.~\cite{Bartsch:2024ofb}.
	
	\paragraph*{\bf Leading single-soft coefficient}---After the order-$\tau^0$ Adler-zero term vanishes, the quartic two-derivative formulation gives a simple way to organize the leading coefficient in the tree-level single-soft expansion~\cite{Cachazo:2016njl}.  
	\begin{figure*}[htbp]
		\centering
		\includegraphics[width=0.9\textwidth]{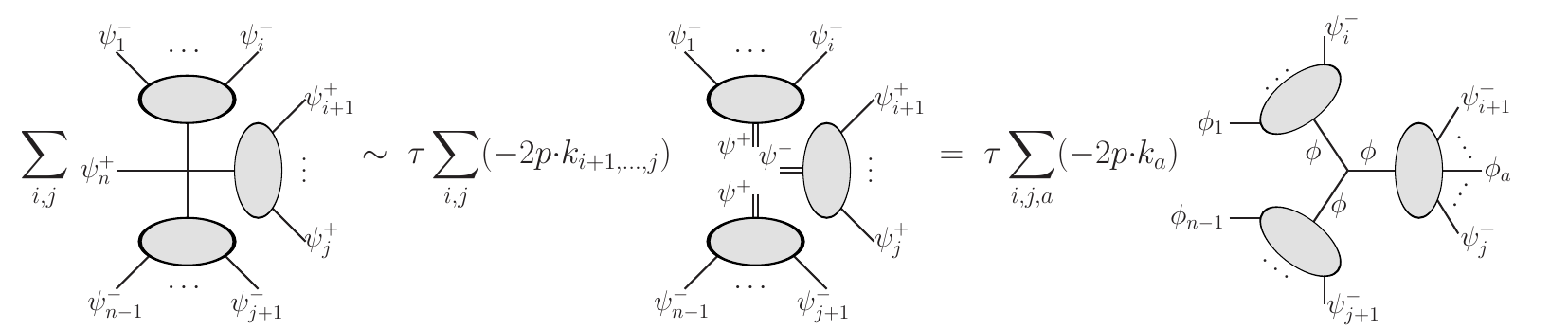}
		\caption{Diagrammatic organization of the leading single-soft coefficient in the QTDS formulation.}
		\label{fig:ssl}
	\end{figure*}
	
	As summarized in Fig.~\ref{fig:ssl}, after taking $k_n{=}\tau p$, the leading term factorizes into the soft vertex factor $-2\tau p{\cdot}k_{i+1,\ldots,j}$ times three tree blocks; the double lines denote off-shell legs together with their kinematic poles.  Since a pure-pion amplitude is equivalent to the mixed amplitude with only two $\phi$ insertions, the three tree blocks may be represented as mixed blocks.  These blocks are then glued by a cubic $\phi^3$ vertex, with legs $1$ and $n{-}1$ fixed as $\phi$ legs and the third scalar $\phi_a$ inserted at all possible positions in the remaining block. Expanding $k_{i+1,\ldots,j}$ as $\sum_{i<a\leq j}k_a$ distributes the soft factor over the possible positions of $\phi_a$, giving
	\begin{equation}
		A_n^\psi|_{k_n\to\tau p}=\tau\sum_{a{=}2}^{n{-}2}(-2p{\cdot}k_a) A_{n-1}^{\psi{+}\phi}(1,a,n{-}1)+\mathcal{O}(\tau^2)\,.
	\end{equation}
	Here $A_{n-1}^{\psi{+}\phi}(1,a,n{-}1)$ denotes the mixed amplitude with ordering $\phi_1\psi^+_2\ldots\phi_a\ldots\psi^+_{n{-}2}\phi_{n{-}1}$.  Using the mixed-amplitude equivalence, this reproduces the leading single-soft coefficient of Ref.~\cite{Cachazo:2016njl}.
	
	In recent work~\cite{Arkani-Hamed:2023swr,Bartsch:2024amu,Li:2024bwq}, the authors identified hidden zeros for a class of theories including the NLSM: tree amplitudes vanish on certain special kinematic loci.  In particular, the Adler zero follows from a class of hidden zeros referred to there as skinny zeros.  The quartic rule makes another class of zero loci manifest, which appears to be special to the NLSM.  Indeed, each quartic diagram vanishes on either of the loci $s_{ee'}{=}0$ for all even labels $e,e'$, or $s_{oo'}{=}0$ for all odd labels $o,o'$.

	\subsection{Double-soft behaviors}
	Double-soft limits contain the next layer of universal infrared data of the NLSM~\cite{DuLuo:2016MultiSoft,Arkani-Hamed:2024fyd}.  While the single-soft limit vanishes by the Adler zero, taking two pions soft can produce a nontrivial factorization onto a lower-point amplitude.  The result depends on the relative positions of the two soft legs in the color ordering.  In this subsection we focus on the leading double-soft behavior.  For adjacent soft legs, the simultaneous soft limit has a finite leading term governed by the standard double-soft factor associated with the commutator of broken generators~\cite{Cachazo:2015ksa,Low:2015ogb}.  For next-to-adjacent soft legs, this finite term vanishes, and the first nonzero contribution is linear in the soft parameter~\cite{Du:2015esa}.  We show that these leading contributions follow directly from the local quartic rule of QTDS: only a small set of diagrams contains the required soft pole, and those diagrams reproduce the corresponding soft factors.
	
	\paragraph*{\bf Double-soft factor for adjacent legs}---We consider a $(n{+}2)$-point pion amplitude $A^\pi_{n+2}$ with $n\geq4$, so that the first nontrivial case is six points.  The two adjacent soft legs are taken to be $k_{n{+}1}{=}\tau p$ and $k_{n{+}2}{=}\tau q$, with $\tau{\to}0$.  The adjacent double-soft theorem~\cite{Kampf:2013vha, Cachazo:2015ksa} states that $A^\pi_{n+2}|_{k_{n{+}1}\to\tau p,\,k_{n{+}2}\to\tau q}=S^{(0)} A^\pi_n+\mathcal{O}(\tau)$, where the leading soft factor is
	\begin{equation}
		S^{(0)}
		=\frac{1}{2}\bigg[
		\frac{k_n{\cdot}(p{-}q){+}\tau\,p{\cdot}q}
		{k_n{\cdot}(p{+}q){+}\tau\,p{\cdot}q}
		{+}\frac{k_1{\cdot}(q{-}p){+}\tau\,p{\cdot}q}
		{k_1{\cdot}(q{+}p){+}\tau\,p{\cdot}q}\bigg]
		\label{eq:double-soft-S0}
	\end{equation}
	
\begin{figure}[htbp]
		\centering
		\includegraphics[width=0.98\linewidth]{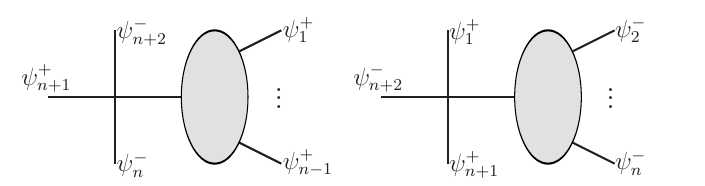}
		\captionsetup{justification=raggedright,singlelinecheck=false}
		\caption{The two local configurations that contribute at leading order in the adjacent double-soft limit.  They contain the soft poles $s_{n,n{+}1,n{+}2}$ and $s_{n{+}1,n{+}2,1}$, both of order $\mathcal{O}(\tau)$, and can be viewed as gluings of a four-point vertex to an $n$-point current.}
		\label{fig:dsfadj}
	\end{figure}

	We now evaluate the same soft limit in QTDS.  As shown in Fig.~\ref{fig:dsfadj}, the quartic rule~\eqref{eq:quarticrule} restricts the leading contribution to two local configurations.
	Evaluating these two configurations gives
	\begin{equation}
		\begin{aligned}
			&A^\psi_{n{+}2}|_{k_{n{+}1}\to\tau p,\,k_{n{+}2}\to\tau q}\\
			=&
			\frac{2k_{n{+}1}{\cdot} k_{n,n{+}1,n{+}2}}{s_{n,n{+}1,n{+}2}}
			A^\psi_{n}|_{k_n\to k_n{+}\tau(p{+}q)}+
			\\
			&\frac{-2k_{1}{\cdot} k_{n{+}1}}{s_{n{+}1,n{+}2,1}}
			A^\psi_{n}|_{k_1\to k_1{+}\tau(p{+}q)}+\mathcal{O}(\tau)
			\\
			=&\left[\frac{p{\cdot}(k_n{+}\tau q)}{k_n{\cdot}(p{+}q){+}\tau\,p{\cdot}q}{-}\frac{k_1{\cdot}p}{k_1{\cdot}(p{+}q){+}\tau\,p{\cdot}q}\right]A^\psi_n{+}\mathcal{O}(\tau)
			\\
			=&S^{(0)}\,A^\psi_n{+}\mathcal{O}(\tau)
		\end{aligned}
	\end{equation}
	where in the second equality we used $A^\psi_{n}|_{k_a\to k_a{+}\tau p}=A^\psi_n{+}\mathcal{O}(\tau)$ and the on-shell condition $p^2=0$.

\begin{figure}[htbp]
		\centering
		\includegraphics[width=0.55\linewidth]{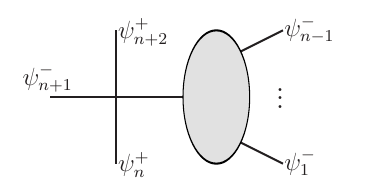}
		\captionsetup{justification=raggedright,singlelinecheck=false}
		\caption{The local configurations that contribute at leading order in the next-to-adjacent double-soft limit. }
		\label{fig:dsfnadj}
	\end{figure}

	\paragraph*{\bf Double-soft factor for next-to-adjacent legs}---The quartic rule~\eqref{eq:quarticrule} also makes the leading next-to-adjacent behavior transparent.  In this limit the two soft legs are separated by one hard leg and share a common hard neighbor.  We take $k_{n}{=}\tau p$ and $k_{n{+}2}{=}\tau q$, with $\tau{\to}0$, and choose the polarity assignment in which even legs carry positive polarities.  The finite leading term is absent.  Compared with the adjacent double-soft limit, the only allowed soft pole is now $s_{n,n{+}1,n{+}2}$, so the first nonzero contribution comes only from the local configuration shown in Fig.~\ref{fig:dsfnadj}.
	
	Similarly, the resulting leading term is
	\begin{equation}
		A^\psi_{n{+}2}|_{k_{n}\to\tau p,\,k_{n{+}2}\to\tau q}=\tau \frac{-p{\cdot} q}{k_{n{+}1}{\cdot}(p{+}q)}A^\psi_n+\mathcal{O}(\tau^2)\,,
		\label{eq:dsfnadj}
	\end{equation}
	which agrees with Ref.~\cite{Du:2015esa}. For the next-to-adjacent configuration the situation parallels the Adler zero: a fixed polarity assignment makes only one half of the possible next-to-adjacent double-soft limits manifest, while the globally flipped assignment makes the complementary half manifest. 
	
	For all other relative positions of the two soft legs, no quartic diagram contains a propagator that becomes soft, and the leading term starts at order $\tau^2$.
	
	\paragraph*{\bf At loop level}---For the QTDS integrand representatives, the same local soft-pole argument extends naturally on generalized cuts.  The singular soft poles used above can only come from the local tree-like part of a diagram, where the two soft legs are attached to the adjacent hard data through a quartic vertex and a soft propagator.  The rest of the graph may contain an arbitrary loop subdiagram, but it is seen by the soft factor only through the corresponding lower-point integrand.  Thus, on generalized cuts, the double-soft factor multiplies the all-loop lower-point QTDS integrand in the same way as at tree level. 
	This is an all-loop realization, in the QTDS formulation, of the planar flavor-ordered form of the leading double-soft pion theorem.  At the amplitude level, the same leading soft factor is fixed by current algebra, as reviewed for example in Ref.~\cite{BereanDutcher:2025hsy}.

	\section{Conclusion and Outlook}
	\label{sec:outlook}
	
	We have shown that planar NLSM amplitudes admit a local quartic two-derivative formulation. The infinite tower of contact interactions is replaced by two auxiliary scalar fields and a single quartic two-derivative vertex, reproducing amplitudes at all-loop level on generalized cuts.
	
	The formulation makes the Adler zero and the leading double-soft factor transparent at all loop orders on generalized cuts.  The two global polarity assignments manifest complementary Adler zeros, while adjacent and next-to-adjacent double-soft limits reduce to simple soft-pole configurations.  It also gives a compact tree-level organization of the leading single-soft coefficient and provides a more economical expansion for high-multiplicity amplitudes.
	We also explored this construction for mixed $\rm NLSM{+}\phi^3$ amplitudes. Low-point amplitudes determine natural local $\psi$--$\phi$ quartic vertices, whereas higher multiplicities require additional contact terms, including a penta-vertex for certain polarity patterns. Whether a finite set of local vertices suffices remains an open question.
	
	Several further questions are left for future work. First, the Lagrangian origin of QTDS should be clarified, including its relation to the standard NLSM, its flavor-dressed formulation, and its connection to shifted-kinematics descriptions~\cite{Arkani-Hamed:2024nhp}. Second, the quartic rules may provide an efficient framework for studying subleading soft behavior. More broadly, it is natural to ask whether similar finite-valency reorganizations exist for other EFTs with enhanced soft behavior, such as the special Galileon~\cite{Cachazo:2014xea}. Despite its multiplicity-dependent derivative interactions in $D$-dimension, the special Galileon, as the double copy of two NLSMs, provides a natural testing ground for these ideas.
	
	\section*{Acknowledgements}
	We thank for comprehensive discussions with Song He. The work of Q.C. is supported by the Westlake Fellows Program at Westlake University. 
	The work of F.Z. is supported in part by the Science Challenge Project (No. TZ2025012), and NSAF No. U2330401.
	
	\bibliographystyle{apsrev4-1}
	\bibliography{Refs}
	
	\clearpage
	
	\appendix
	\onecolumngrid
	
	\section{Proof for equivalence between NLSM and QTDS amplitudes}
	\label{app:proof}
	
	In this appendix we give the diagrammatic proof that the quartic QTDS rule reproduces the planar NLSM amplitudes.  The tree-level proof is most transparent in planar variables.  The key point is that factorization reduces the comparison to a boundary term, equivalently to a local contact term.  After this contact term is matched, generalized unitarity extends the same conclusion to loop amplitudes.  We close the appendix by explaining how the quartic rules make the algebraic soft limits~\cite{Bartsch:2024ofb} manifest for half of the external legs.
	
	\paragraph*{\bf Planar variables}----For a fixed cyclic canonical ordering $\mathbb{I}_{2n}{=}(1,2,\ldots,2n)$, we introduce the planar variables with labels understood cyclically
	\begin{equation}
		X_{i,j}=(x_j-x_i)^2=\left(p_i+p_{i+1}+{\cdots}+p_{j-1}\right)^2\,,
		\label{eq:planarX}
	\end{equation}
	where $x_i$ is the dual coordinate.  For external massless particles $X_{i,i+1}{=}p_i^2{=}0$.  A chord $(i,j)$ in the dual polygon represents the planar channel $X_{i,j}$.  
	
\begin{figure*}[htbp]
		\centering
		\includegraphics[scale=0.6]{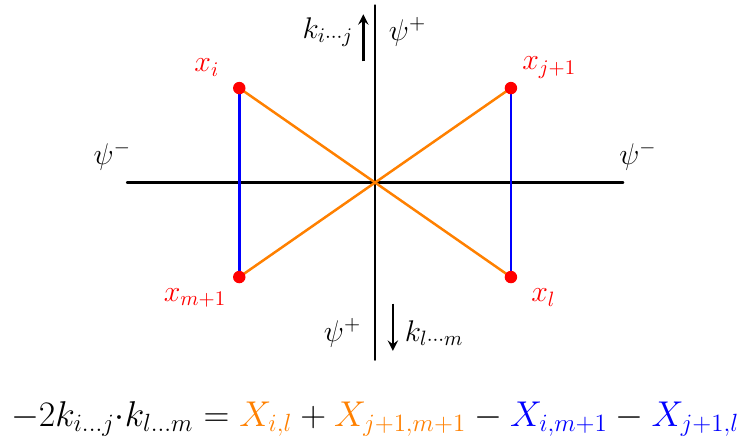}
		\captionsetup{justification=raggedright,singlelinecheck=false}
		\caption{The quartic rule in dual coordinates.  The red points denote the dual coordinates surrounding the quartic vertex.  In planar variables, each vertex is written as a difference between two crossed chords and two parallel chords.}
		\label{fig:Xrule}
	\end{figure*}

	The QTDS vertex admits a simple translation into these variables.  When the four corners of the quartic vertex are related by dual points, the scalar product in the momentum-space rule becomes a difference of the two crossed chords and the two parallel chords, as shown in Fig.~\ref{fig:Xrule}.
	
\begin{figure*}[htbp]
		\centering
		\includegraphics[scale=0.65]{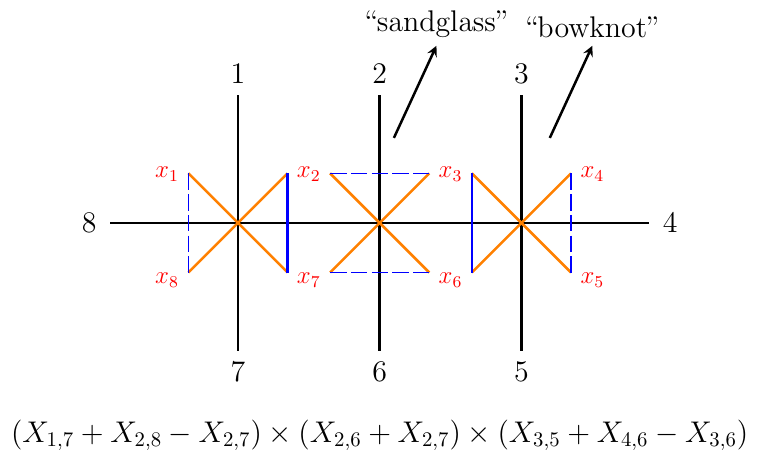}
		\captionsetup{justification=raggedright,singlelinecheck=false}
		\caption{Example of the numerator for an eight-point quartic diagram.  The ``bowknot'' and ``sandglass'' icons denote contributions from single quartic vertices.  Dashed lines indicate variables that vanish, such as $X_{2,3}=p_2^2=0$.}
		\label{fig:8ptexp}
	\end{figure*}

	Figure~\ref{fig:8ptexp} shows an eight-point example.  We will use the two local patterns in this figure, the ``sandglass'' and the ``bowknot'', to track which planar variables appear in each vertex. The important {\it combinatorial constraint} is that two adjacent icons cannot be of the same type: adjacent vertices cannot both be sandglasses, and cannot both be bowknots.

	We compare the QTDS result with the NLSM amplitude as rational functions of the planar variables.  On any physical pole, the residue factorizes into a product of lower-point amplitudes.  Assuming the equality has already been proved at lower multiplicity, all physical residues of the two representations agree.  Their difference therefore has no physical poles.  At tree level such a pole-free difference is precisely the boundary term left invisible to factorization, namely a local polynomial contact term.  Since the two rules agree at four points, the induction is complete once the $2n$-point contact projection of the quartic rule is shown to reproduce the minimal NLSM contact vertex~\eqref{eq:catalan-rule}.
	
	\paragraph*{\bf Contact projection}----
	In the minimal parametrization, the $2n$-point contact term can be written in planar variables as
	\begin{equation}
		V_{\pi}^{(2n)}
		=
		\frac{(-1)^n}{2}
		\sum_{k=1}^{n-1} C_{k-1}C_{n-k-1}
		\sum_{i=1}^{2n}X_{i,i+2k}\,,
		\label{eq:catalan-rule}
	\end{equation}
	where $C_k{=}\frac{1}{k+1}\binom{2k}{k}$ is the Catalan number. This is the ``Catalan representation'' of the minimal NLSM contact vertices in Ref.~\cite{Arkani-Hamed:2024yvu}.  Since every on-shell NLSM subamplitude has even multiplicity, physical propagators always correspond to opposite-parity variables, $X_{e,o}$ or $X_{o,e}$.  In contrast, the polynomial contact data are naturally organized by same-parity variables, $X_{e,e}$ and $X_{o,o}$, as shown in Eq.~\eqref{eq:catalan-rule}.
	
	We now show that the quartic rule reconstructs exactly the same contact data.
	
	Consider a single $2n$-point quartic tree.  It contains $n{-}1$ quartic vertices and $n{-}2$ propagators.  To extract its contact contribution, one must choose $n{-}2$ planar variables from the numerator to cancel all propagators.  The remaining vertex then supplies the local contact term.  After deleting the dashed blue lines that correspond to vanishing planar variables, each remaining blue line corresponds to a unique propagator.  Since different vertices do not share the same propagator variable, this cancellation can occur only in the following form:
	\begin{quote}
		\it exactly $n{-}2$ selected vertices each contain one blue line, and each such blue line cancels one propagator, while the remaining vertex contains no blue line and contributes to the $2n$-point contact term.
	\end{quote}
	These restrictions leave a very special class of quartic diagrams.  
	\begin{figure*}[htbp]
		\centering
		\includegraphics[scale=0.65]{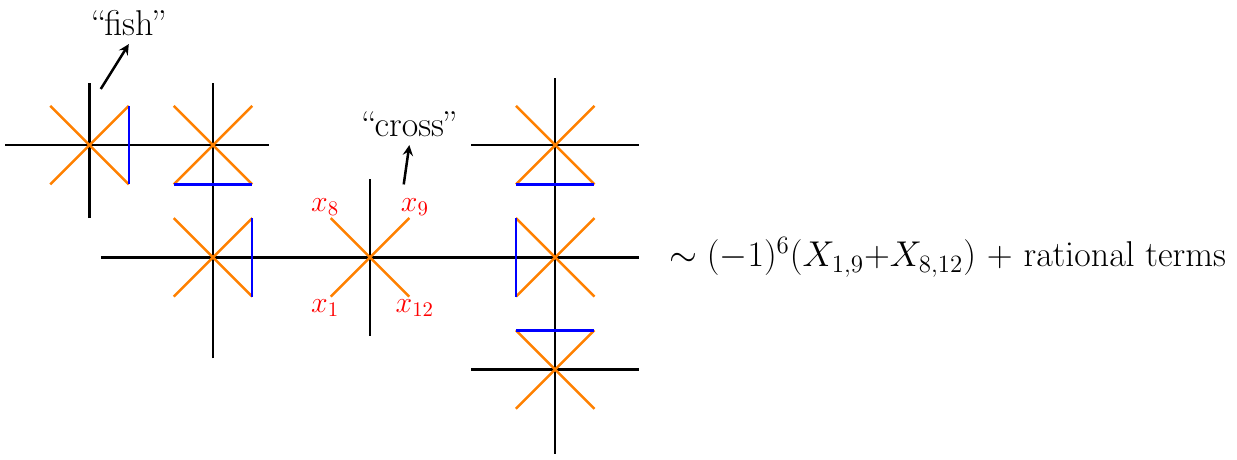}
		\captionsetup{justification=raggedright,singlelinecheck=false}
		\caption{A 16-point quartic diagram contributing to the contact term.  The central ``cross'' supplies the contact variable, while each ``fish'' supplies a blue line that cancels a propagator and contributes an additional factor of $-1$.}
		\label{fig:fish-app}
	\end{figure*}
	As shown in Fig.~\ref{fig:fish-app}, a selected vertex can be represented by a ``fish'' icon, while the remaining vertex is represented by a ``cross'' icon.  All fishes must be oriented so that they can reach the cross by following propagators.  In addition, no pair of adjacent fishes can point in the same direction.  Once the cross is fixed, the allowed selections therefore split into two independent distributions, one on each side of the cross.
	
\begin{figure}[htbp]
		\centering
		\includegraphics[scale=0.75]{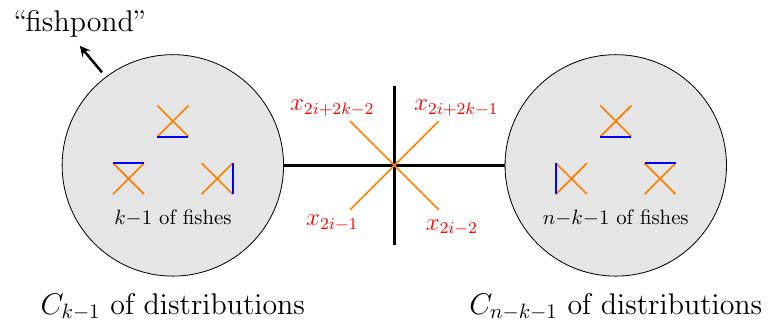}
		\captionsetup{justification=raggedright,singlelinecheck=false}
		\caption{A fishpond represents the collection of possible distributions of fishes on one side of a fixed cross.  The multiplicity of $X_{2i{-}1,2i{+}2k{-}1}{+}X_{2i{-}2,2i{+}2k{-}2}$ is the product of the two fishpond counts, $C_{k-1}C_{n-k-1}$.}
		\label{fig_fishpond}
	\end{figure}

	Consider the cross that contributes $X_{2i-1,2i+2k-1}{+}X_{2i-2,2i+2k-2}$. 
	The two sides of this cross contain $k{-}1$ and $n{-}k{-}1$ fishes.  We call the collection of possible fish distributions on one side a ``fishpond''.  The multiplicity of the cross contribution is therefore the product of the two fishpond multiplicities, see Fig.~\ref{fig_fishpond}.
	
\begin{figure*}[htbp]
		\centering
		\includegraphics[scale=0.8]{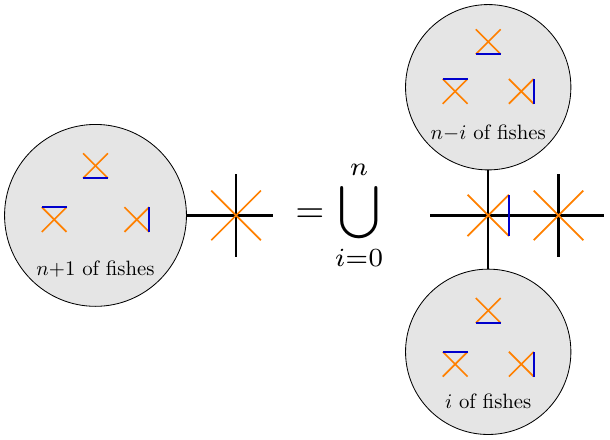}
		\captionsetup{justification=raggedright,singlelinecheck=false}
		\caption{Recursive distributions of a fishpond. The counting is given by the recursion $C_{n+1}=\sum_{i=0}^n C_iC_{n-i}$ with offset $C_0=C_1=1$.}
		\label{fig:catalanfish}
	\end{figure*}

	It remains to count the allowed distributions of a fishpond.  The counting follows the same recursion as the Catalan numbers.  As shown in Fig.~\ref{fig:catalanfish}, a fishpond with $n{+}1$ fishes decomposes into two smaller fishponds on the two sides of the fish closest to the central cross.  Therefore the number of allowed distributions satisfies the Catalan recursion.
	
	Collecting all possible crosses, the contact projection of the quartic rule is
	\begin{equation}
		\begin{aligned}
			V_\psi^{(2n)}
			&=
			\frac{(-1)^{n{-}2}}{2}
			\sum_{i=1}^{n}\sum_{k=1}^{n-1}
			C_{k-1}C_{n-k-1}
			\left(
			X_{2i-1,2i+2k-1}
			+X_{2i-2,2i+2k-2}
			\right)
			\\
			&=
			\frac{(-1)^n}{2}
			\sum_{k=1}^{n-1}
			C_{k-1}C_{n-k-1}
			\sum_{i=1}^{2n}X_{i,i+2k}.
		\end{aligned}
		\label{eq:quartic-contact-original-app}
	\end{equation}
	The second equality is just a reorganization of the cyclic sum over $i$.  As $k$ runs from $1$ to $n{-}1$, the complementary chords are counted twice, and the overall factor $1/2$ removes this overcounting.  Equation~\eqref{eq:quartic-contact-original-app} is exactly the Catalan contact rule~\eqref{eq:catalan-rule}.  Together with the factorization argument above, this proves the tree-level equivalence between the QTDS and NLSM planar amplitudes. The globally flipped assignment is treated in the same way.  The individual quartic diagrams are not generally equal to their flipped counterparts, but the contact projection obeys the same Catalan counting and reproduces the same contact rule~\eqref{eq:catalan-rule}.

	Following is the list of tree-diagram counts used in the efficiency discussion in Sec.~\ref{sec:quartic}.
	\begin{table*}[htbp]
		\centering
		\scriptsize
		\setlength{\tabcolsep}{4pt}
		\renewcommand{\arraystretch}{1.65}
		\begin{tabular}{@{}cccc@{}}
			\hline\hline
			& Quartic trees (\href{https://oeis.org/A001764}{OEIS:A001764}) & Cubic trees (Catalan number) & Even-valent trees (\href{https://oeis.org/A003168}{OEIS:A003168}) \\ 
			\hline
			Counting for $(2k{+}4)$-point&
			$\displaystyle
			\frac{1}{k+1}\binom{3k{+}3}{k}$
			&
			$\displaystyle
			\frac{1}{2(k{+}1)}\binom{4k{+}4}{2k{+}1}$
			&
			$\displaystyle
			\begin{gathered}
				\frac{2^{k{-}1}}{k}\binom{3k}{k{-}1}{}_2F_1(-2k{-}1,-2k;-3k;{-}1)
			\end{gathered}$
			\\[5pt]
			Asymptotic growth &
			$\displaystyle \sim\sqrt{\frac{3}{16\pi}}\frac{(27/4)^{k{+}1}}{(k{+}1)^{3/2}} $
			&
			$\displaystyle \sim \frac{16^{k{+}1}}{\sqrt{\pi(2k{+}2)}(2k{+}3)}$
			&
			$\displaystyle \sim 
			\sqrt{\frac{14+66/\sqrt{17}}{\pi }}\frac{\left(17 \sqrt{17}+71\right)^k}{2^{4 (k+1)}k^{3/2}}$
			\\[5pt]
			First 5 terms &
			$3,\,12,\,55,\,273,\,1428$
			&
			$14,\,132,\,1430,\,16796,\,208012$
			&
			$4,\,64,\,1344,\,32256,\,837632$
			\\
			\hline\hline
		\end{tabular}
		\captionsetup{justification=raggedright,singlelinecheck=false}
		\caption{Diagram counting for quartic, cubic, and even-valent trees.  Here ${}_2F_1$ is the Gauss hypergeometric function.}
		\label{tab:qtds-counting}
	\end{table*}
	
	\paragraph*{\bf Loop-level}----
	The extension to loop amplitudes follows from generalized unitarity~\cite{Bern:1994zx,Britto:2004GenUnitarity,BernHuang:2011GenUnitarity}.  A planar generalized cut puts a set of internal propagators on shell and decomposes the loop integrand into products of tree amplitudes.  In the QTDS formulation, the alternating polarity assignment is inherited consistently by the cut internal lines, so every tree factor appearing on the cut is one of the amplitudes whose equality was proved above.  Hence every generalized cut of the QTDS integrand agrees with the corresponding planar NLSM cut.
	
	This statement should be understood as an equality of amplitudes, or equivalently of integrand representatives modulo the usual generalized-unitarity ambiguity.  Terms that have no generalized cuts, including scaleless integrands in dimensional regularization, integrate to zero and are not fixed by the cut construction.  Thus the quartic QTDS rules reproduce the all-loop planar NLSM amplitudes, while different polarity choices may correspond to different but cut-equivalent integrand representatives.
	
	At tree level, the equality of the two global polarity choices followed from the Catalan contact-term argument above.  At loop level, however, the ``fishpond'' configuration of Fig.~\ref{fig_fishpond} is no longer preserved, so this argument does not directly extend to loop integrands.  Starting at two loops, opposite polarity patterns can therefore give different integrand representatives, even though their difference is cut-free and integrates to zero.
	
	\begin{figure*}[htbp]
		\centering
		\includegraphics[scale=0.65]{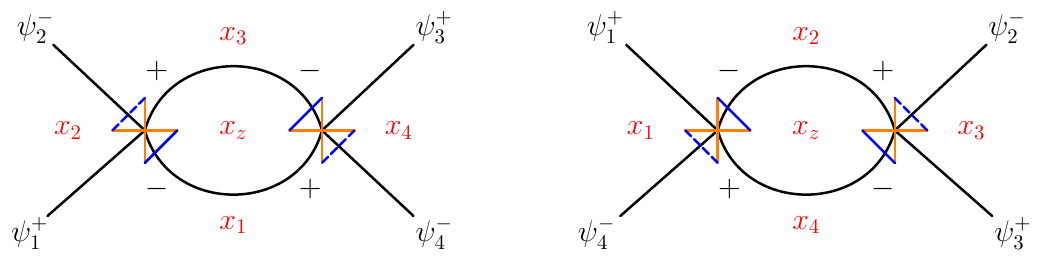}
		\captionsetup{justification=raggedright,singlelinecheck=false,hypcap=false}
		\caption{The one-loop four-point example.  Here $x_z$ is the dual-coordinate associated with the loop momentum $\ell$, and we define $X_{i,z}=(\ell{+}k_{1,\ldots,i{-}1})^2$.  Tadpole-like diagrams are omitted since they integrate to zero.  The quartic rule gives $\frac{(X_{1,3}{+}X_{2,z}{-}X_{1,z})(X_{1,3}{+}X_{4,z}{-}X_{3,z})}{X_{1,z}X_{3,z}}
			+\frac{(X_{2,4}{+}X_{1,z}{-}X_{2,z})(X_{2,4}{+}X_{3,z}{-}X_{4,z})}{X_{2,z}X_{4,z}}$.}
		\label{fig:1loop4pt-app}
	\end{figure*}
	
	\paragraph*{\bf Algebraic soft limits}----
	It is also useful to phrase the Adler zero directly in planar variables.  Following Ref.~\cite{Bartsch:2024ofb}, we define the algebraic soft limit of leg $i$ by replacing the label $i$ by label $i{+}1$ in planar variables, which is the dual-coordinate degeneration $x_i\to x_{i+1}$, and then $p_i^\mu=(x_i-x_{i+1})^\mu\to0$.
	
\begin{figure*}[htbp]
		\centering
		\includegraphics[scale=0.75]{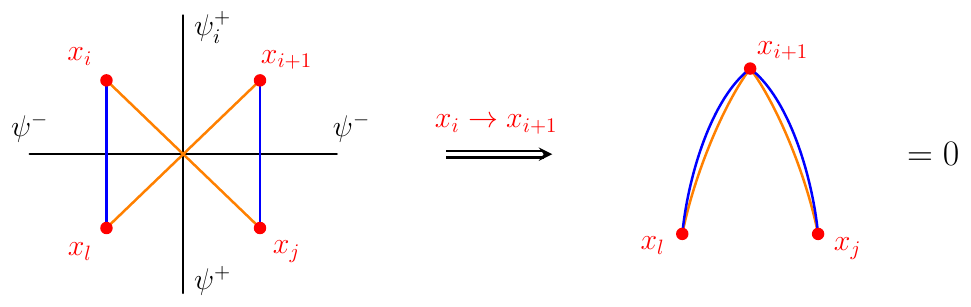}
		\captionsetup{justification=raggedright,singlelinecheck=false}
		\caption{The algebraic soft limit of $\psi^+_i$.  The dual coordinates are identified, $x_i\to x_{i+1}$, so that the local vertex vanishes.}
		\label{fig:algrbraicST}
	\end{figure*}

	For a fixed polarity assignment, the QTDS vertex adjacent to the soft leg then degenerates locally, as shown in Fig.~\ref{fig:algrbraicST}.  The two pairs of chord contributions become identical and cancel.  Thus this algebraic form makes the soft zero manifest diagram by diagram for one half of the external legs; the flipped polarity assignment makes the other half manifest.
	
	\onecolumngrid
	\section{Mixed Amplitudes and Polarity Constraints}
	\label{app:mixed}
	
	In this appendix we collect simple examples of mixed amplitudes.  These examples illustrate why the polarity assignments in the mixed sector are constrained, rather than arbitrary.
	
\begin{figure}[htbp]
		\centering
		\includegraphics[scale=0.7]{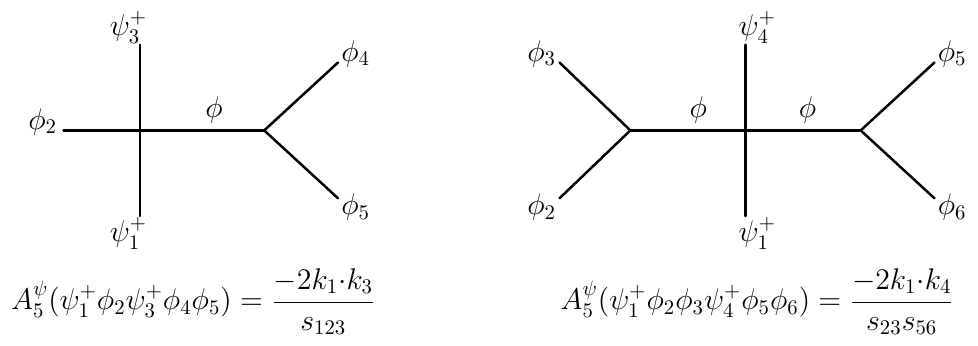}
		\caption{The mixed examples at five- and six- points.}
		\label{fig:56mixexp}
	\end{figure}

	Figure~\ref{fig:56mixexp} shows two simple QTDS$+\phi^3$ examples, each of which receives contribution from a single Feynman diagram.  These examples are also useful for seeing how the penta-vertex in Eq.~\eqref{eq:pentavertex} enters after a global flip of the $\psi$ polarities: the flipped assignment is no longer reproduced by the quartic diagram alone, and requires an additional five-point contact contribution.
	
\begin{figure}[htbp]
		\centering
		\begin{subfigure}[t]{0.32\linewidth}
			\centering
			\includegraphics[width=\linewidth]{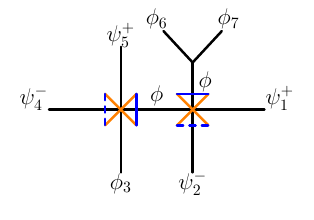}
			\caption{}
		\end{subfigure}
		\hfill
		\begin{subfigure}[t]{0.32\linewidth}
			\centering
			\includegraphics[width=\linewidth]{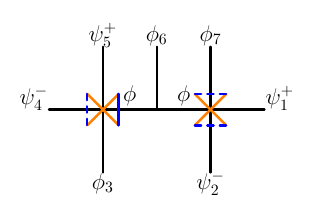}
			\caption{}
		\end{subfigure}
		\hfill
		\begin{subfigure}[t]{0.32\linewidth}
			\centering
			\includegraphics[width=\linewidth]{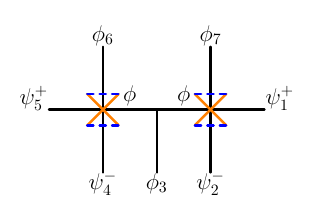}
			\caption{}
		\end{subfigure}
		
		\vspace{0.35cm}
		
		\begin{subfigure}[t]{0.32\linewidth}
			\centering
			\includegraphics[width=\linewidth]{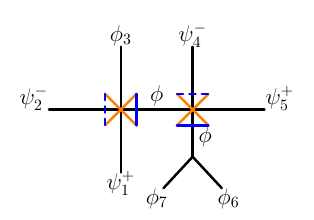}
			\caption{}
		\end{subfigure}
		\hfill
		\begin{subfigure}[t]{0.32\linewidth}
			\centering
			\includegraphics[width=\linewidth]{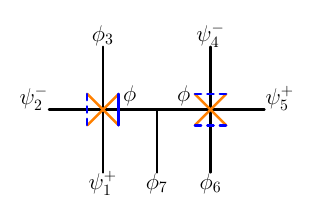}
			\caption{}
		\end{subfigure}
		\hfill
		\begin{subfigure}[t]{0.32\linewidth}
			\centering
			\includegraphics[width=\linewidth]{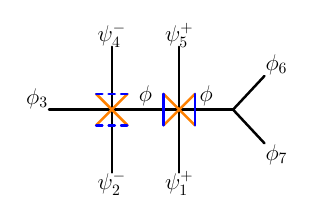}
			\caption{}
		\end{subfigure}
		\caption{The six diagrams contributing to the seven-point mixed amplitude $A^\psi_7(\psi_1^+\psi_2^-\phi_3\psi_4^-\psi_5^+\phi_6\phi_7)$.}
		\label{fig:7mixexp}
	\end{figure}

	Figure~\ref{fig:7mixexp} shows a nontrivial example for the ordering $\psi_1^+\psi_2^-\phi_3\psi_4^-\psi_5^+\phi_6\phi_7$.  We again use the ``sandglass'' and ``bowknot'' icons of Fig.~\ref{fig:8ptexp} to denote the local quartic vertices.  If instead one chooses the polarity pattern $\psi_1^+\psi_2^-\phi_3\psi_4^+\psi_5^-\phi_6\phi_7$, then in the diagrams corresponding to (a), (d), and (f) in Fig.~\ref{fig:7mixexp}, two adjacent vertices would be represented by icons of the same type.  As explained in Appendix~\ref{app:proof}, such adjacent equal icons are incompatible with the contact-term structure needed to recover the higher-point NLSM vertex $V_\pi^{(2m)}$.  This is why the alternative polarity pattern is not allowed. 
	
	The first example with more than two $\psi$-blocks of length greater than one appears at nine points, with the schematic ordering $$\psi\,\psi\,\phi\,\psi\,\psi\,\phi\,\psi\,\psi\,\phi\,.$$ For this distribution, any choice of polarity pattern encounters the same obstruction described above: some adjacent vertices are necessarily associated with the same icon.  Thus amplitudes of this type cannot be recovered by a simple extension of the polarity assignment alone, and require new rules or additional local input.
	
	We now record a few further representative orderings and their allowed polarity constraints.  Different patterns give rise to different sets of contributing Feynman diagrams, and the vertex types appearing in these diagrams may also vary; in particular, some patterns require the penta-vertex in Eq.~\eqref{eq:pentavertex}, as illustrated in Fig.~\ref{fig:mix-flip-example}.  Let $\Psi_m$ denote a consecutive $\psi$-block of length $m$, and let $\Phi$ denote a nonempty consecutive sequence of $\phi$ fields.  Different occurrences of $\Psi_m$ or $\Phi$ may represent different $\psi$- or $\phi$-runs of the indicated type.  The polarities within every nontrivial $\psi$-block continue to alternate.
	\begin{itemize}[leftmargin=2em,itemsep=0.15em,topsep=0.2em,parsep=0pt]
		\item 
		$\Psi_m\Phi\Psi_1\Phi\Psi_1$ with $m{\geq}2$: the polarity choices of the three $\psi$-blocks are independent.  This gives the maximal number of independent binary choices for this type of ordering.
		
		\item
		$\Psi_m\Phi\Psi_2\Phi\Psi_1\Phi$ with $m{\geq}2$: the boundary $\psi$ fields of $\Psi_2$ and $\Psi_1$ that face $\Psi_m$ carry the same polarity.
		
		\item
		$\Psi_m\Phi\Psi_1\Phi\Psi_1\Phi\Psi_1\Phi$ with $m{\geq}2$: the first and last $\Psi_1$ blocks carry the same polarity as the boundary $\psi$ field of $\Psi_m$ that they face, while the polarity of the middle $\Psi_1$ block is independent.
		
		\item
		$\Psi_2\Phi\Psi_2\Phi\Psi_1\Phi\Psi_1\Phi$: the boundary $\psi$ fields of the two $\Psi_2$ blocks that face one another carry the same polarity, while the polarity choices of the two $\Psi_1$ blocks are independent.
	\end{itemize}

\end{document}